\documentclass[preprint,aps,showpacs]{revtex4}
\usepackage{graphicx}
\def\nablaperp{\nabla_{\!\!\perp}}
\def\Real{\mathop{\rm Re}\nolimits}
\def\Imag{\mathop{\rm Im}\nolimits}
\newcommand{\eqn}[1]{(\ref{#1})}
\newcommand{\eqns}[2]{(\ref{#1})-(\ref{#2})}
\newcommand{\fig}[1]{Fig.~\ref{#1}}

\def\Vec#1{\mbox{\boldmath $#1$}}
\def\d{{\rm d}}
\def\mathfrac{\frac{\mathstrut}{\mathstrut}}
\newcommand{\lsim}{\mbox{\
\raisebox{-.6ex}{$\stackrel{\textstyle<}{\sim}$}\ }}
\newcommand{\gsim}{\mbox{\
\raisebox{-.6ex}{$\stackrel{\textstyle>}{\sim}$}\ }}

\begin{document}

\title{Angular momenta creation in relativistic electron-positron plasma}
\author{T. Tatsuno}
\affiliation{Graduate School of Frontier Sciences, The University of Tokyo,
	Tokyo 113-0033, Japan}
\author{V. I. Berezhiani}
\affiliation{Institute of Physics, The Georgian Academy of Sciences,
	Tbilisi 380077, Georgia}
\author{M. Pekker and S. M. Mahajan}
\affiliation{Institute for Fusion Studies, The University of Texas at Austin,
	Austin, Texas 78712}

\date{\today}

\begin{abstract}
Creation of angular momentum in a relativistic electron-positron plasma is explored.
It is shown that a chain of angular momentum carrying vortices is a robust asymptotic state sustained by the generalized nonlinear Schr\"odinger equation characteristic to the system.
The results may suggest a possible electromagnetic origin of angular momenta when it is applied to the MeV epoch of the early Universe.
\end{abstract}


\maketitle

\section{Introduction}

The problem of electromagnetic (EM) wave propagation and related phenomena
in relativistic plasmas has attracted considerable attention in the recent
past. From the nonthermal emission of the high-energy radiation coming from
a variety of compact astrophysical objects it has become possible to deduce
the presence of a population of relativistic electrons in the plasma created
in the dense radiation fields of those sources \cite{1}. The principal components
of these plasmas could be either relativistic electrons and nonrelativistic
ions (protons), or relativistic electron-positron (e-p) pairs.

Relativistic e-p dominated plasmas may be created in a variety of astrophysical situations.
The e-p plasmas are likely to be found in pulsar magnetospheres \cite{pulsar}, in the bipolar outflows (jets) in Active Galactic Nuclei (AGN) \cite{AGN}, and at the center of our own Galaxy \cite{Galaxy}.
The presence of e-p plasma is also argued in the MeV epoch of the early Universe.
In the standard cosmological model, temperatures in the MeV range ($T\sim 10^ {10}$K${}- 1$MeV) prevail up to times $t =1$sec after the Big Bang \cite{universe}.
In this epoch, the main constituent of the Universe is the relativistic e-p plasma in equilibrium with photons, neutrinos, and antineutrinos.

Contemporary progress in the development of super strong laser pulses with
intensities $I\sim 10^{21-23} {\rm W/cm}^{2}$ has also made it possible to create relativistic plasmas in the laboratory by a host of experimental techniques \cite{laser1}.
At the focus of an ultrastrong laser pulses, the electrons can acquire
velocities close to the speed of light opening the possibility of simulating
in the laboratory the conditions and phenomena that, generally, belong in
the astrophysical realm \cite{laser2}.

Elucidation of the electromagnetic wave dynamics in a relativistic plasmas
will, perhaps, be an essential tool for understanding the radiation
properties of astrophysical objects as well as of the media exposed to the
field of superstrong laser radiation. Although the study of wave propagation
in relativistic plasmas has been in vogue for some time, it is only in the
recent years that the nonlinear dynamics of EM radiation in e-p dominated
plasmas \cite{8} has come into focus. The enhanced interest stems from two facts: 1) e-p plasmas seem to be essential constituents of the universe, and 2)
under certain conditions, even an ultrarelativistic electron-proton plasma
can behave akin to an e-p plasma \cite{9}.

Recently, we firstly found dark soliton as well as vortex soliton solutions 
in e-p plasma  \cite{TBM-01}.
In Ref.~\cite{FaBu-dark}, it is also shown analytically that dark soliton is the natural nonlinear
coherent structure in unmagnetized cold e-p plasmas.
However, it is conceivable that soliton solutions obtained in a one dimensional
 formulation will turn out to be unstable in higher dimensions, which may lead to the creation of vortex solitons.
Since dark and vortex solitons are asymptotically nonvanishing, they have received much less attention than their localized cousins due to the generally accepted requirement that the fields be localized in physical system.
However, in recent experiments studying laser field dynamics in different kinds of optical media, it was demonstrated that dark and vortex solitons can be readily created as superimpositions upon a localized field background \cite{KL-98}.
Vortex soliton solutions are also found in imperfect Bose gas in the superfluids \cite{Pitaevskii}, and they are extensively investigated 
and discussed \cite{BEcondense}.
In this paper, we systematically investigate the instability of dark solitons and show that it can lead to the formation of vortex solitons.
We speculate about interesting application of the vortex soliton in the early universe.


In the recent paper \cite{TBM-01}, we had developed an argument for the creation of domains of nonzero angular momentum in the MeV era of the early universe when it is supposed to be dominated by a plasma of e-p pairs.
We first showed that in such a plasma the dynamics of a pulse of electromagnetic radiation, with a frequency much larger than the plasma frequency, is controlled by a generalized nonlinear Schr\"odinger equation (GNSE) with a defocusing nonlinearity.
Then, borrowing a result of nonlinear optics \cite{KL-98} where the standard nonlinear Schr\"odinger equation (NSE) with a cubic nonlinearity has been investigated in great depth and detail, we conjectured that even the GNSE whose nonlinearity is similar in nature to that of NSE will allow vortex soliton solutions with an angular momentum that is conserved during propagation.
The latter system also allows dark soliton solutions in one dimension which are known to be unstable to two-dimensional (2d) perturbations and eventually evolve to 2d vortex chains.

In this paper we demonstrate that the arguments given in Ref.~\cite{TBM-01} are actually borne out by the direct solutions of the derived GNSE.
We begin by analytically showing that the 1d dark solitons of the GNSE are, indeed, unstable to 2d perturbations.
In the process, we derive the instability criterion which agrees with the numerical solution of the linearized system.

The main part of the demonstration, however, comes from a numerical simulation of our GNSE.
Starting from a broad variety of initial conditions, we find the eventual emergence of angular momentum carrying vortex solitons.
That is, the stipulated solitonic structures are readily accessible within the framework of this general equation with a nonlinearity more complicated than that of NSE.
Thus we can state with much greater confidence that electromagnetism, operative in the MeV era, could easily be the primordial source of angular momentum associated with various structures of the observable universe.

We again stress that the present results are quite general in the nonlinear dynamics in an electron positron plasma.

\section{Single vortex soliton}
\label{single vortex soliton}

In the envelope approximation, a finite amplitude, circularly polarized electromagnetic (EM) pulse propagating in a relativistic e-p plasma obeys (for details, see \cite{TBM-01}):
\begin{equation}
	2 {\rm i} \omega \partial_{\tau} A_{\perp} + c^2\nabla_{\!\!\perp}^{2}
		A_{\perp} + \frac{c^2}{\gamma_{g}^{2}} \partial_{\xi}^{2} A_{\perp}
		{} + \frac{2 \omega_e^2}{G_0} \biggl( 1 - \frac{G_0}{\gamma G}
		\frac{n}{n_0} \biggr) A_{\perp} = 0,
\label{NSE}
\end{equation}
where $A_{\perp}$ is the slowly varying amplitude of the perpendicular (to the
propagation direction) vector potential, $\nablaperp^2 = \partial_x^2 + \partial_y^2$, and $\xi = z - v_g t$ is the ``comoving'' (with group velocity $v_g$) coordinate.
Here $\omega_e=(4\pi n_{0e} e^2 / m_e)^{1/2}$ is the electron Langmuir frequency, $n_0 = n_{0e} = n_{0p}$ is the number density of the unperturbed background (subscript $0$ denotes the value at infinity), $\gamma$ is the Lorentz factor, and $G = K_{3}(m_e c^2/T)/K_{2}(m_e c^2/T)$ with $K_{n}$, denoting the $n\/$-th order modified Bessel functions of the second kind, is the enthalpy density.

In this approximation the continuity equation, determining the number density $n$, becomes
\begin{equation}
\frac{1}{\gamma G} \frac{n}{n_0} = \frac{v_g \gamma_g / c}{[
		\gamma_g^2 G_0^2 - G^2 - e^2|A_{\perp}|^2/(m_e c^2)^2]^{1/2}},
\label{plasma current}
\end{equation}
and the system is closed by the adiabatic equation of state
\begin{equation}
	\frac{n m_e c^2 / n_0 T}{\gamma K_{2}(m_e c^2 / T)}
		\exp (- m_e c^2 G / T) = \mbox{\rm const.}
\label{eq.state}
\end{equation}

The low frequency motion of the plasma is driven by the ponderomotive pressure [${} \sim (\mbox{\boldmath $p$}_{e,p})^{2}$] of the high-frequency EM field, and is independent of the sign of the particles' charge.
It is perfectly natural to assume that the electron and the positron fluids have equal temperatures ($T_{0e,p} = T_{0}$) in equilibrium so that their effective masses ($G_{e,p} = G$) will also be equal.
The radiation pressure will impart equal low-frequency momenta to both fluids allowing the possibility of overall density changes without producing charge separation.
The charge neutrality conditions $n_e = n_p = n$, $\phi = 0$ has been assumed by neglecting the small inequality of the charge due to baryon asymmetry.
It is also evident that the symmetry between the two fluids keeps their temperatures always equal ($T_{e,p}=T$) if they were equal initially.
In deriving \eqns{NSE}{plasma current}, we have also assumed that the plasma is transparent (i.e. $\omega \gg \omega_e$), and that the longitudinal extent of the pulse is much shorter than its transverse dimensions ($L_{\parallel} \ll L_{\perp}$).

Defining the normalized variables
\begin{equation}
	\frac{\omega_e^2}{2 \omega G_0} t \to t, \quad
	\frac{\omega_e}{c \sqrt{2G_0}} \Vec{r} \to \Vec{r}, \quad
	\frac{e}{m_e c^2} \frac{1}{\gamma_g G_0} A_{\perp} \to A_{\perp},
\end{equation}
and noting that even with the assumption $\partial_{\xi} \gg \nablaperp$, the diffractive term can be the same order or even greater than the dispersive one for a highly transparent plasma ($\gamma_g \gg 1$), Eq.~(\ref{NSE}) converts to the following GNSE ($\gamma_g G_0 \gg G$):
\begin{equation}
	{\rm i} \partial_t A + \frac{1}{2} \nablaperp^2 A + f(|A|^2) A = 0,
\label{final NSE}
\end{equation}
with the nonlinearity
\begin{equation}
	f(|A|^2) = -2 \left( \frac{1}{\sqrt{1 - |A|^2}} - 1 \right).
\label{sec2:nonlinearity}
\end{equation}

Equation~(\ref{final NSE}) admits a symmetric two-dimensional
solitary wave solution.
For stationary solitons, the ansatz
\begin{equation}
	A = A(r) \exp ( {\rm i} m \theta - {\rm i} \lambda t),
\end{equation}
reduces \eqn{final NSE} to
\begin{equation}
	\frac{{\rm d}^2}{{\rm d}r^2} A + V'(A) = - \frac{1}{r}
		\frac{{\rm d} A} {{\rm d} r} + \frac{m^2}{r^2} A,
\label{2D-Newtonian}
\end{equation}
with the potential $V(A)$ defined as
\begin{equation}
	V(A) = ( \lambda + 2 ) A^2 + 4 \sqrt{1 - A^2} - 4,
\label{potential representation}
\end{equation}
with the prime denoting the derivative with respect to its argument, and $\lambda$ representing the nonlinear frequency shift.

A numerical analysis of the solution of this eigenvalue problem is given in Ref.~\cite{TBM-01}.
Because of the absence of $\xi$-derivatives in \eqn{final NSE}, any pulse-like localized structure is allowed in the propagation direction.
In the direction perpendicular to the propagation, on the other hand, the amplitude was shown to approach a constant value.
Since there does exist a small baryon asymmetry in the early Universe that acts on a longer characteristic length~\cite{BM-95,MBM-98}, the extent of the constant amplitude region may be considered to be finite.

Due to the single-valuedness of the vector potential, $m$ must be an integer and $A$ must vanish at the origin for non-zero values of $m$.
The nonzero $m$ solutions are particularly important because they carry the orbital angular momentum $\Vec{M}$
\begin{equation}
	(\Vec{M})_z = \frac{{\rm i}}{2} \int {\rm d} \Vec{r}_{\perp} \,
		[\Vec{r}_{\perp} \times (A^* \nablaperp A - {\rm c.c.})]_z.
\label{angular momentum}
\end{equation}
It is straightforward to show that the system \eqn{final NSE} conserves angular momentum, and the expression \eqn{angular momentum} is just the paraxial approximation for the orbital angular momentum, $\Vec{M}_{\!\! E} = \int {\rm d} \Vec{r} \, [ \Vec{r} \times (\Vec{E} \times \Vec{B}) ]$, of the EM field \cite{AB-92}.
The angular momentum carried by the vortex is $M_z = mN$ where the photon number $N =\int {\rm d} \Vec{r}_{\perp} \, |A|^2$ is another conserved quantity; $m$ is also known as the ``topological charge.''
Strictly speaking, one must redefine the integrals of motion for non-vanishing boundary conditions \cite{perturbation theory}, but such a renormalization is not important here because of the fact that infinite-extent solution is just a formal approximation.
The presence of a small fraction of ions makes the physical solution decay at infinity \cite{BM-95,MBM-98}.

\section{Dynamics of angular momenta creation}

It is already suggested in Ref.~\cite{TBM-01} that dark stripe solitons are unstable and break into vortex filaments.
When the amplitude is small ($|A| \ll 1$), the nonlinearity reduces to the simple Kerr-type, and the vortex dynamics of NSE with such a nonlinearity has been studied in a variety of numerical calculations \cite{KL-98,perturbation theory,McDonald,Iv-97}.
Our full nonlinearity is somewhat different and it has to be independently investigated.
For this section, our goal is to show that the one-dimensional dark stripe solution is unstable against transverse perturbations in two dimensions.
Such an instability causes the breakup of the stripe leading to a chain of vortex solitons with alternating polarity.

\subsection{Transverse instability}

The one-dimensional stationary solution for the GNSE (\ref{final NSE}) has the form $g(x) e^{-{\rm i} \lambda t}$ ($\lambda > 0$), where the real function $g(x)$ satisfies
\begin{equation}
	\lambda g + \frac{1}{2} \frac{\d^2 g}{\d x^2} + f(g^2) g = 0.
\label{IIIA:stationary state}
\end{equation}
Since we are examining the linear stability of the solution with the least number of nodes, $g$ is assumed to be an odd function with a single node at the origin.

Let us perturb the one-dimensional solution by 
\begin{equation}
	A = (g + u + {\rm i} v) e^{-{\rm i} \lambda t},
\label{IIIA:linearization}
\end{equation}
where the small perturbations $u$ and $v$ are real functions of $x$ and $y$.
 
Assuming sinusoidal behavior, $u \propto \cos (ky - {\mit \Omega} t)$ and $v \propto \sin (ky - {\mit \Omega} t)$, the linearized GNSE could be reduced to the following eigenvalue problem:
\begin{equation}
	{\mit \Omega}^2 v = {\cal L}_1 {\cal L}_0 v - \frac{k^2}{2} ({\cal L}_0
	+ {\cal L}_1) v + \frac{k^4}{4} v.
\label{combined full spectral ODE for v}
\end{equation}
where
\begin{eqnarray}
	{\cal L}_0 &=& \lambda + \frac{1}{2} \frac{\partial^2}{\partial x^2}
		+ f(g^2), \\
	{\cal L}_1 &=& \lambda + \frac{1}{2} \frac{\partial^2}{\partial x^2}
		+ f(g^2) + 2 g^2 f'(g^2).
\end{eqnarray}

Remembering that ${\cal L}_0 g = 0$ and ${\cal L}_1 g_x = 0$, we can construct from \eqn{combined full spectral ODE for v} the relation
\begin{equation}
	{\mit \Omega}^2 \langle g_x \, | \, v \rangle = \frac{k^4}{4} \langle g_x\,
	| \, v \rangle + \langle g_x \, | \, {\cal L}_1 {\cal L}_0 v \rangle
	- \frac{k^2}{2} \langle g_x \, | \, 2 {\cal L}_1 v \rangle
	- \frac{k^2}{2} \langle g_x \, | \, -2g^2 f' v \rangle
\label{integrated full combined ODE for v}
\end{equation}
by multiplying $g_x$ on both sides and integrating with respect to $x$.
Self-adjointness of ${\cal L}_1$ reduces \eqn{integrated full combined ODE for v} to
\begin{equation}
	{\mit \Omega}^2 = \frac{k^4}{4}-\frac{k^2}{2} \frac{\langle g_x \, | \,
	-2g^2 f'v \rangle}{\langle g_x \,|\, v \rangle},
\label{exact growth rate by integrals}
\end{equation}
provided $\langle g_x \, | \, v \rangle$ exists.
A necessary condition for an exponential instability, then, is
\begin{equation}
	\frac{\langle g_x \, | \, -2g^2 f' v \rangle}{\langle g_x \, | \, v
	\rangle} > 0.
\label{instability condition}
\end{equation}
If it is true, then for sufficiently small $|k|$, we would have an instability, i.e., \eqn{instability condition} is also a sufficient condition.
Moreover, it is explicitly shown that $\mit \omega \to 0$ in the limit $k \to 0$ unless $\langle g_x \, | \, v \rangle$ diverges.

We now make an approximate estimate of the ratio on the left hand side of \eqn{instability condition}.
Let us assume that the operators ${\cal L}_0$ and ${\cal L}_1$ are of order unity and ${\mit \Omega}^2$ and $k^2$ are, in some sense, small.
In this limit, we could estimate $v$ by solving
\begin{equation}
	{\cal L}_1 {\cal L}_0 v = 0
\end{equation}
which has the solution $v = g + q$, where
\begin{equation}
	{\cal L}_0 q = g_x.\label{definition of q}
\end{equation}
Notice that for $v=g$, the integrals occurring in \eqn{exact growth rate by integrals} are zero and this part will not contribute to the integral; only $q$ will.
From \eqn{definition of q}, we can deduce the following:
the first consequence
\begin{eqnarray}
	\langle g_x \, | \, g_x \rangle &=& \langle g_x \, | \, {\cal L}_0 q
	\rangle= \langle g_x \, | \, ({\cal L}_0 - {\cal L}_1) q \rangle\nonumber\\
	&=& \langle g_x \, | \, -2g^2 f' q \rangle = \langle g_x \, | \, -2 g^2
	f'v \rangle
\end{eqnarray}
converts \eqn{exact growth rate by integrals} into
\begin{equation}
	{\mit \Omega}^2 = \frac{k^4}{4} - \frac{k^2}{2} \frac{\langle g_x \,|\,
	g_x \rangle}{\langle g_x \, | \, v \rangle}.
\end{equation}
To derive the second, we notice that
\begin{equation}
	\langle g_x \, | \, v \rangle = \langle g_x \, | \, q \rangle
	= \langle g_x \, | \, {\cal L}_0^{-1} g_x \rangle
\label{denominator of dispersion}
\end{equation}
In principle, one can evaluate \eqn{denominator of dispersion} in a rather straightforward way.
But a very approximate estimate can be made by simply dropping $\partial_x^2$ in ${\cal L}_0$ so that
\begin{equation}
	\langle g_x \, | \, v \rangle \simeq \left\langle \mathfrac g_x \,
	\right|\, \left. \frac{g_x}{\lambda + f} \right\rangle
\end{equation}
and
\begin{equation}
	{\mit \Omega}^2 = \frac{k^4}{4} - \frac{k^2}{2} \frac{\langle g_x \,
	| \, g_x \rangle}{\langle g_x \, | \, g_x / (\lambda + f) \rangle}
	\equiv \frac{k^4}{4} - \alpha k^2.
\label{definition of alpha}
\end{equation}
In \eqn{definition of alpha} we have overestimated the denominator.
The eigenvalue is
\begin{equation}
	{\mit \Omega} = {\rm i} \alpha k \left[ 1 - \frac{k^2}{4 \alpha}
		\right]^{1/2},
\label{final analytic eigenvalue}
\end{equation}
implying a window $0 < k < 2\alpha^{1/2}$ in $k$, where instability is possible .
From the property that the one-dimensional solution is stable to one-dimensional perturbation, the growth rate ${\mit \Omega}$ approaches zero as the wave number $k$ tends to zero.

In order to accomplish the numerical analysis, we introduce a complex-valued function $w = u + {\rm i} v$ in \eqn{IIIA:linearization}.
Then, the linearized equation looks
\begin{equation}
  -{\rm i} \partial_t w = \lambda w + \frac{1}{2} \nablaperp^2 w
    + f(g^2) w + 2 g^2 f'(g^2) \Real(w),
\end{equation}
where the prime denotes derivative with respect to its argument.
Putting $\partial_t = -{\rm i} {\mit \Omega}$ and $\partial_y = {\rm i}k$ yields a linear eigenvalue equation for $w$:
\begin{equation}
  - {\mit \Omega} w = \lambda w + \frac{1}{2} \left( \frac{\d^2 w}
    {\d x^2} - k^2 w \right) + f(g^2) w + 2 g^2 f'(g^2) \Real(w).
\label{linst: spectral equation}
\end{equation}
We have numerically solved \eqn{linst: spectral equation} by the shooting method with boundary conditions $w \to 0$ ($x \to \infty$) and $\d w / \d x = 0$ ($x=0$).
It is clear from \eqn{linst: spectral equation} that the complex conjugate of the eigenfunction is also an eigenfunction.
Namely, if $w_{\rm e}$ is an eigenfunction of \eqn{linst: spectral equation} corresponding to the eigenvalue ${\mit \Omega}_{\rm e}$, then $\bar{w}_{\rm e}$ is also an eigenfunction of \eqn{linst: spectral equation} corresponding to the eigenvalue $\bar{\mit \Omega}_{\rm e}$, where the bar denoting the complex conjugate.

For $x \to + \infty$, $g$ approaches a constant value $g_0$, which can be expressed as
\begin{equation}
        g_0 = \sqrt{1 - \biggl( \frac{2}{\lambda + 2} \biggr)^2},
\end{equation}
where the inhomogeneity due to the potential vanishes.
In this region, $x$ also becomes an ignorable direction, and we can assume $w \propto e^{- \kappa x}$ for the point spectra ($\kappa > 0$).
Plugging it into our eigenvalue problem, we obtain the following ``dispersion relation'' which is applicable for large $x$,
\begin{equation}
        -{\mit \Omega}_{\rm i}^2 = \biggl[ {\mit \Omega}_{\rm r} + \frac{1}{2}
		(\kappa^2 - k^2) \biggr]
	\biggl[ {\mit \Omega}_{\rm r} + \frac{1}{2} (\kappa^2 - k^2)
		+ 2 g_0^2 f'(g_0^2) \biggr],
\end{equation}
where the subscripts r and i denote the real and imaginary part, respectively, and we have used the relation $f(g_0^2) = - \lambda$.
By solving it for $\kappa$, we obtain
\begin{equation}
        \kappa = \sqrt{ k^2 - 2 {\mit \Omega}_{\rm r} - 2 g_0^2 f'(g_0^2)
                \pm 2 \sqrt{g_0^4 [f'(g_0^2)]^2 - {\mit \Omega}_{\rm i}^2}},
\label{asymptotics}
\end{equation}
where $f'(g^2)$ is explicitly shown as
\begin{equation}
        f'(u) = - (1 - u)^{-3/2}.
\end{equation}
The asymptotic argument of $w$ is also determined by
\begin{equation}
	\arg(w) = \arctan \left( \frac{2{\mit \Omega}_{\rm i}}
		{k^2 + 2 {\mit \Omega}_{\rm r} - \kappa^2} \right).
\label{linst: asymptotic argument}
\end{equation}

With these informations, we have solved \eqn{linst: spectral equation} numerically.
First, we fix an arbitrary value on the real part of $w$ at $x=10$.
Given an expectant eigenvalue, the imaginary part of $w$ is calculated from \eqn{linst: asymptotic argument}.
Then, the derivative $\d w / \d x$ is estimated by using \eqn{asymptotics}, and we can
integrate \eqn{linst: spectral equation} from $x=10$ to $x=0$ by the 4-th order Runge-Kutta formula.
The data of $g(x)$ is drawn from the numerical solution of \eqn{IIIA:stationary state}.
At $x=0$, the boundary condition $\d w / \d x = 0$ is checked.
If this boundary condition is not satisfied, we guess the next eigenvalue by the Newton method and carry out the shooting again.

\begin{figure} \begin{center}
  \includegraphics[width=12cm]{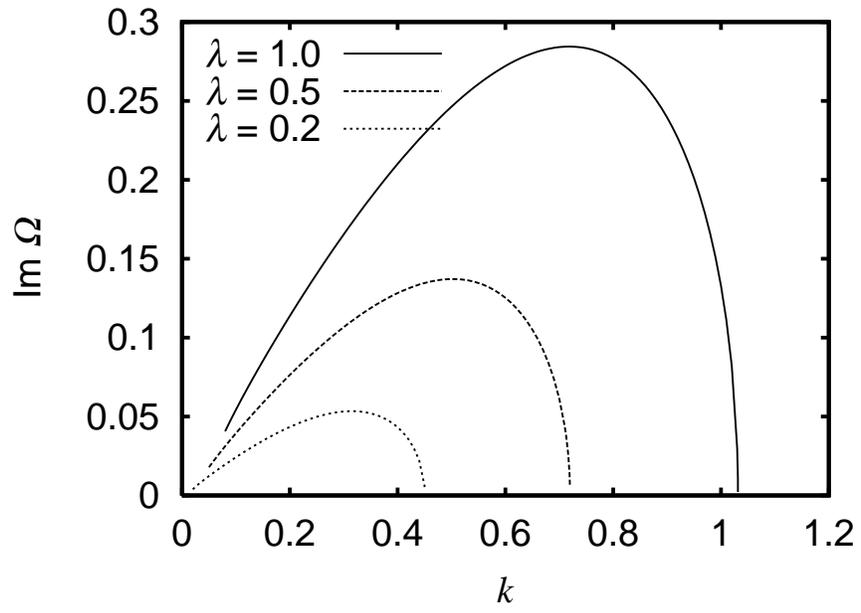}
\end{center}
\caption{Dispersion relation for $\lambda = 0.2$, $0.5$, and $1$.}
\label{fig: dispersion}
\end{figure}

\begin{figure} \begin{center}
  \includegraphics[width=10cm]{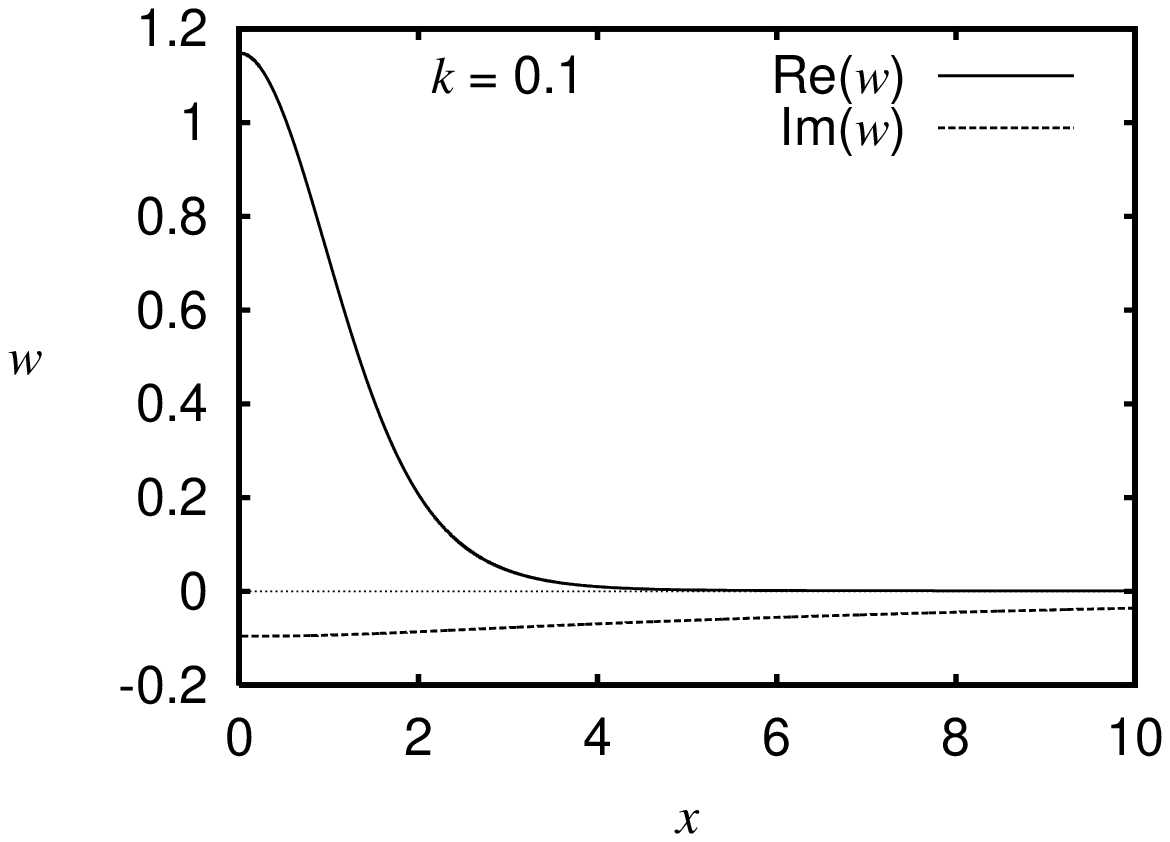}
  \includegraphics[width=10cm]{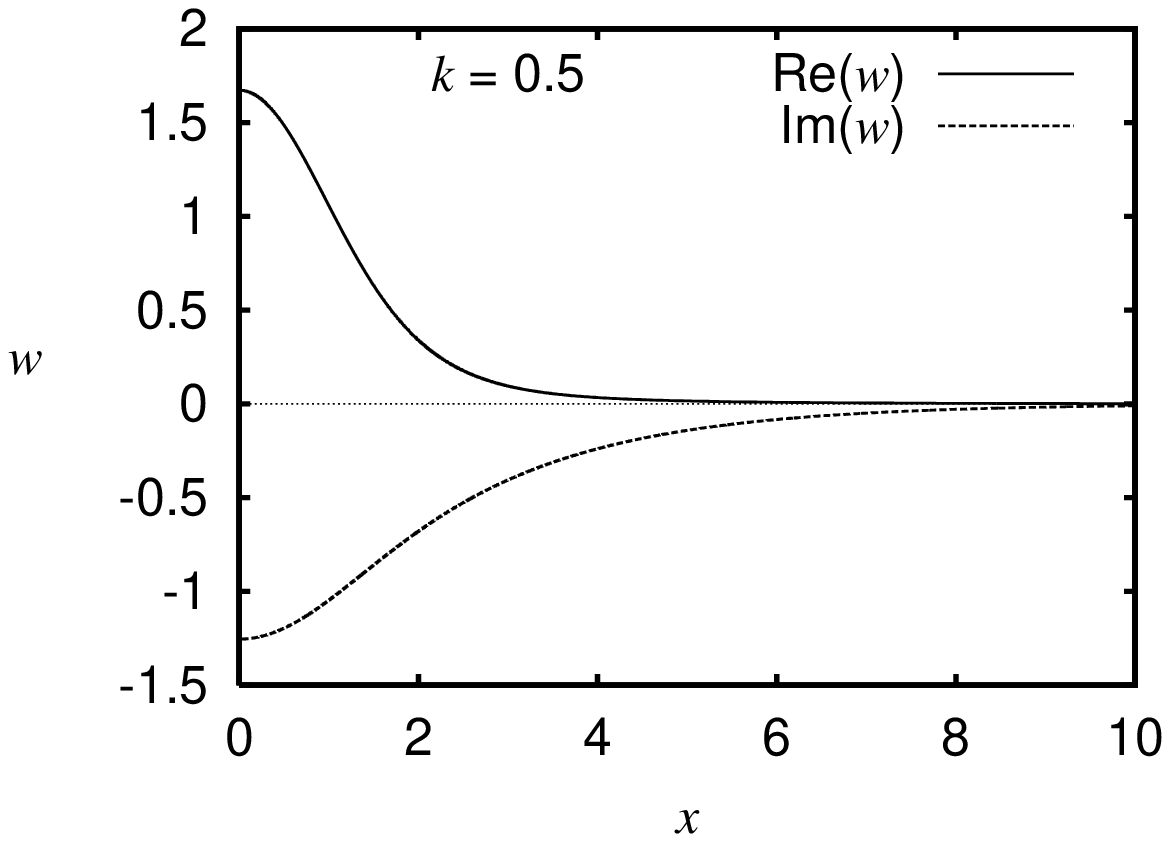}\\
  \includegraphics[width=10cm]{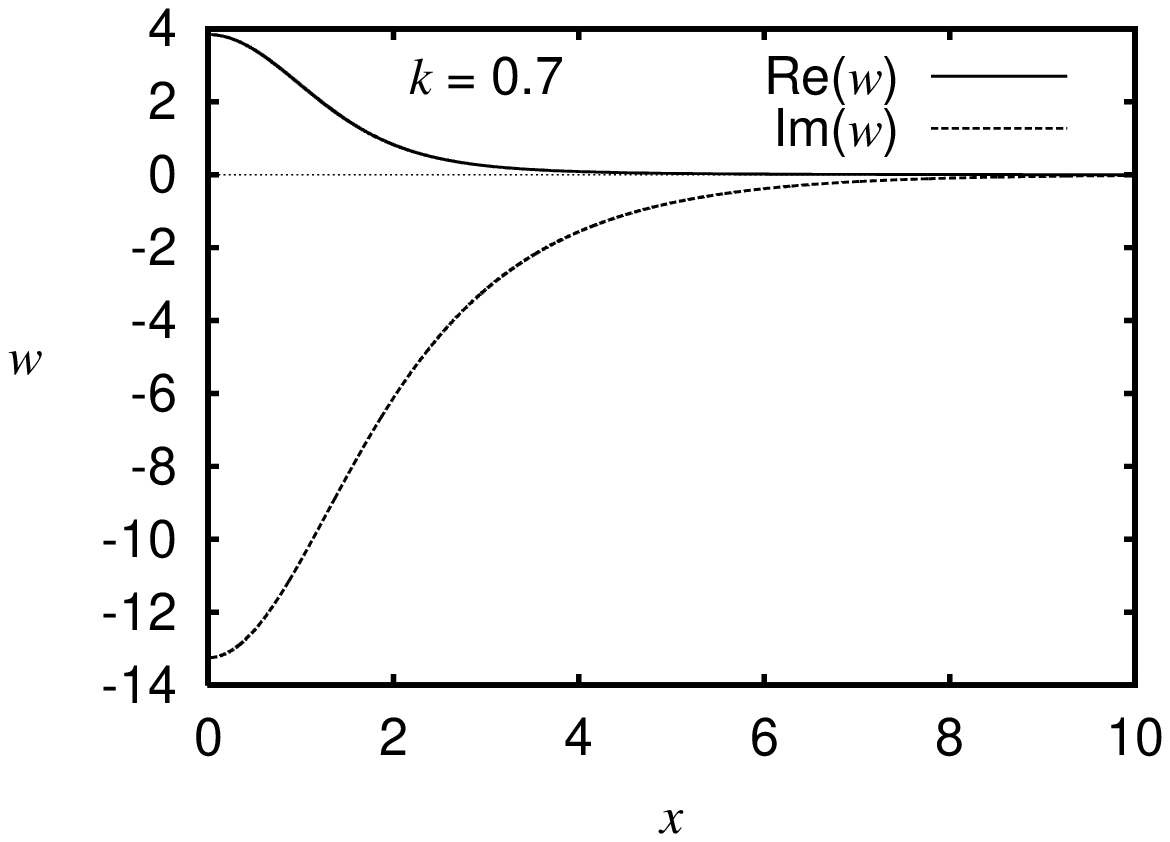}
\end{center}
\caption{Eigenfunctions for $\lambda = 0.5$ corresponding to $k=0.1$, $0.5$, and $0.7$.}
\label{fig: eigenfunctions}
\end{figure}

The dispersion relation is shown in \fig{fig: dispersion}.
We can see a good qualitative agreement with the analytic evaluation with the fact that the growth rate begins from zero as $k \to 0$, experiences a maximum value with respect to $k$, and the mode is finally stabilized for a sufficiently large $k$.
The eigenfunctions corresponding to $k=0.1$, $0.5$, and $0.7$ for the parameter $\lambda = 0.5$ are illustrated in \fig{fig: eigenfunctions}.
From \fig{fig: dispersion}, the mode with $k=0.5$ gives the maximum growth rate $\Imag ({\mit \Omega}) \simeq 0.137$.
As is seen from \fig{fig: eigenfunctions}, the imaginary part of the eigenfunction becomes wider and wider as $k \to 0$, which agrees qualitatively with the analytical estimate.
Since we are performing the numerical shooting in the finite domain, it becomes difficult to estimate the correct eigenvalue in this regime.

\subsection{Nonlinear evolution}

We will now present the numerical simulation of the dynamics of angular momentum creation from one-dimensional dark stripe solitons by solving the full nonlinear system \eqns{final NSE}{sec2:nonlinearity}.
The calculation was carried out on a $200 \times 200$ spatial mesh placed in a calculation box of size $L = 28$.
The boundary conditions $\partial_x A = 0$ ($\partial_y A = 0$) are imposed at the edges $x = \pm L$ ($y = \pm L$).
Note that these boundary conditions numerically respect the constants of motion in the domain of integration.
As an input we choose the one-dimensional $x$ dependent stationary solution with $\lambda = 0.5$.
It is to be noted that the photon number and the Hamiltonian are conserved to order $10^{-5}$ during the calculation.
Since the initial value of the integrated angular momentum is zero, it is found to always remain zero (within an error of order $10^{-15}$).

Strictly speaking, the eigenfunction in the previous section (see \fig{fig: eigenfunctions}) contains cores of angular momenta, i.e. crossings of the real and imaginary zeros of the field due to the periodic form of the perturbation in the $y$ direction.
Thus, the growth of small amplitude perturbation itself does not exactly mean the `creation' of angular momenta.
In order to check the real creation of angular momenta from an exact zero everywhere, we have first carried out the calculation with an initial condition of the form
\begin{equation}
	A(x,y,0) = \alpha \left[ g(x) + \tilde{A}_1 \frac{\d g}{\d x}
		\exp (- \beta y^2) \right] + {\rm i} \sqrt{1 - \alpha^2} g(x),
\end{equation}
with $\alpha = 0.9$, $\beta = 0.0875$, and $\tilde{A}_1 = 0.1$.
This initial condition is so arranged that there is no crossing of the real and imaginary zeros.
It is done by imposing a real-valued perturbation on a complex-valued stationary solution.
The zero line of the imaginary part exactly coincides with the $y$-axis, while that of the real part deviates from the $y$-axis by $\tilde{A}_1$.
As the system evolves, we observe the creation of crossings and the appearance of vortex solitons.
As the initial conditions are changed (always starting form no crossings), the main qualitative result --- the emergence of angular momentum carrying vortex soliton chains --- is found to be fairly robust.

\begin{figure} \begin{center}
	\includegraphics[width=12cm]{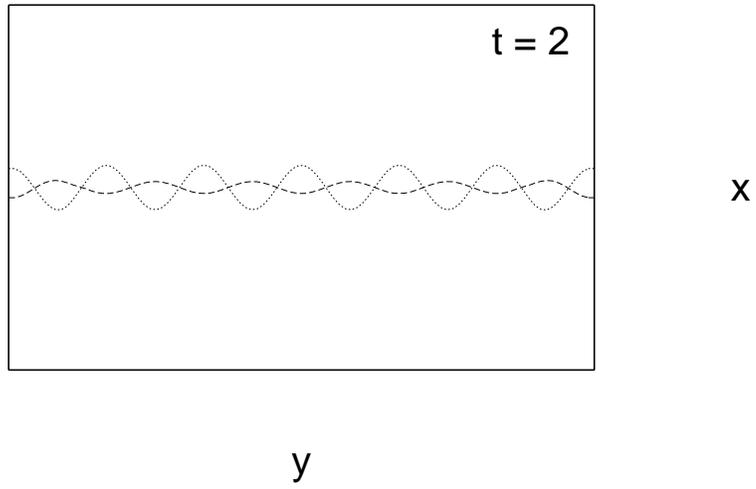}
\end{center}
\caption{Zero lines of real and imaginary parts of the field at time $t = 2$.}
\label{fig:crossing at time 2}
\end{figure}

Next we investigate the evolution of the system under transverse perturbations of the form
\begin{equation}
	A(x,y,0) = g(x) + \tilde{A} \frac{\d g}{\d x}
		\cos \left( \frac{n \pi}{L}y \right),
\label{nonlinear:initial condition 1}
\end{equation}
with $\tilde{A} = 0.1$ and $n = 6$%
.
The zero lines of the real and imaginary parts at $t = 2$ are shown in \fig{fig:crossing at time 2}.
As we noted in Sec.~\ref{single vortex soliton}, the crossing points of two zero lines correspond to the vortex centers.
We initially have twelve crossings suggesting twelve vortices.
These twelve vortices, however, do not have well-formed solitonic structures since they are too close, and overlapping.

As the system evolves, the vortices move and we see annihilations of pairs of opposite polarity.
The first annihilation event was observed in the interval $28 \lsim t \lsim 34$; two of the vortices destroy one another near the center, and the other two seem to disappear near the edge; the annihilation of two pairs removes four from the original twelve at $t=2$.

\begin{figure} \begin{center}
	\includegraphics[width=8cm]{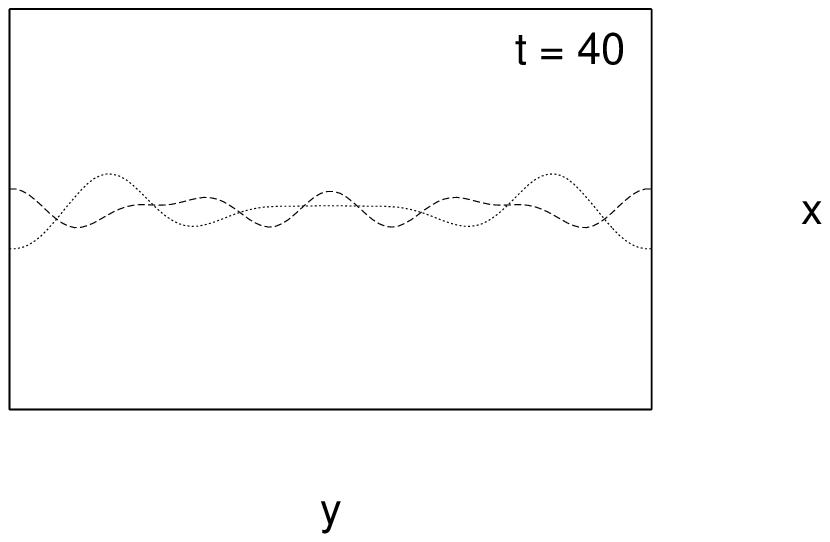}
	\includegraphics[width=8cm]{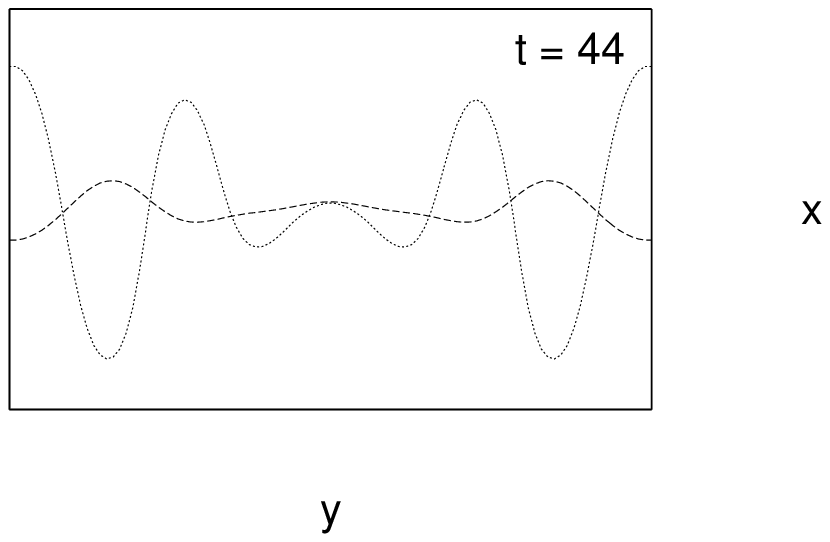}
	\includegraphics[width=8cm]{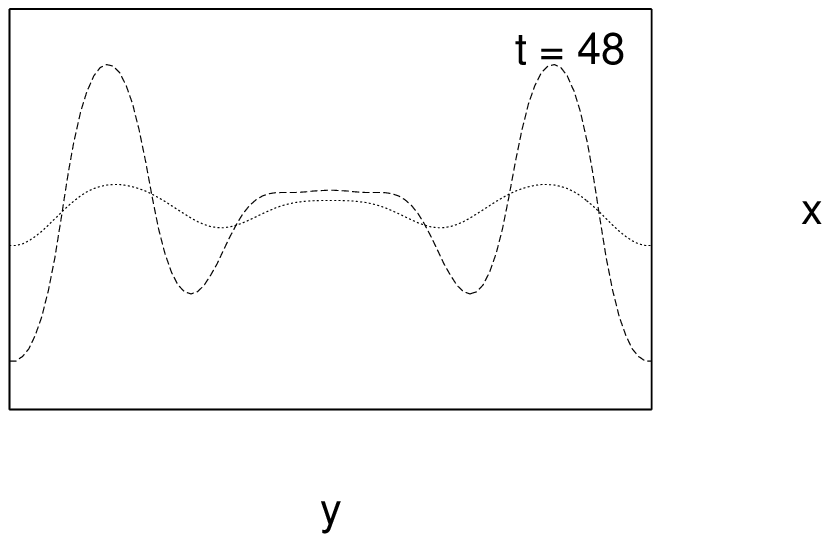}
	\includegraphics[width=8cm]{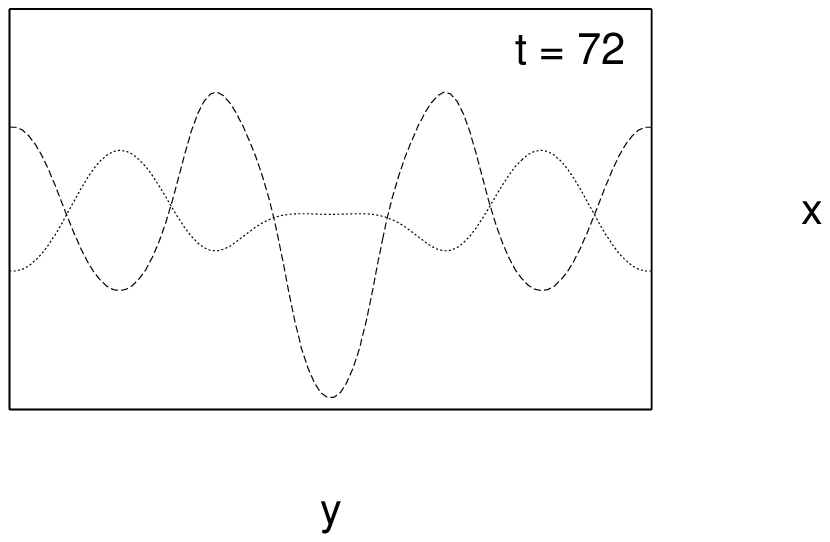}
\end{center}
\caption{Crossing of zeros at time $t = 40$, 44, 48 and 72.}
\label{fig:annihilation}
\end{figure}

The second annihilation event occurs during $40 \lsim t \lsim 48$.
As is depicted in \fig{fig:annihilation}, the zero lines of the real and imaginary parts tend to separate around the center $y \sim 0$.
Here two vortices approach the origin around $40 \lsim t \lsim 44$, and then they are annihilated.
After the annihilation, two inner vortices approach the origin.
Finally the remaining six vortices tend to spread and align with equal inter-vortex spacing (see $t = 72$ in \fig{fig:annihilation}).

\begin{figure} \begin{center}
	\includegraphics[width=8cm]{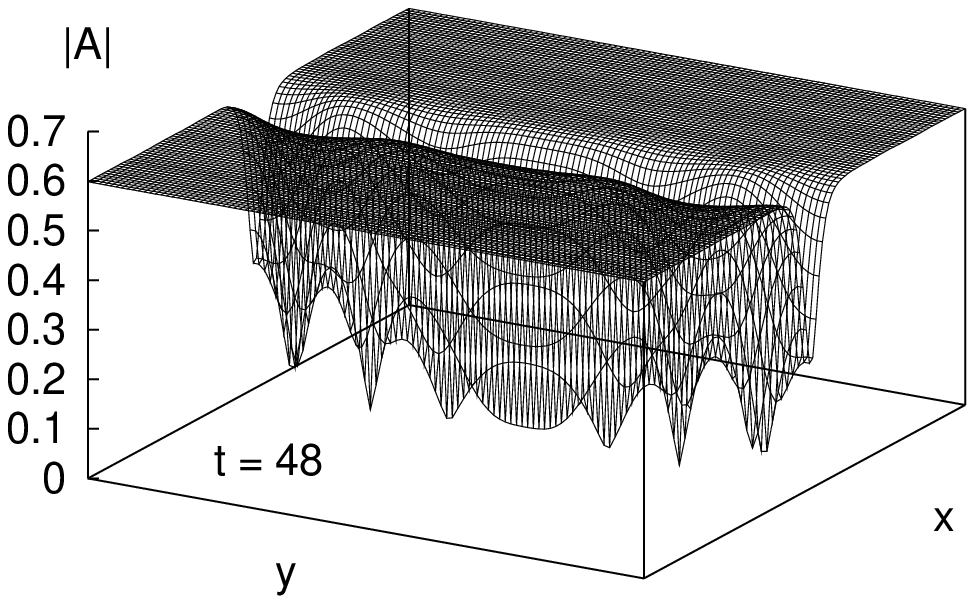}
	\includegraphics[width=8cm]{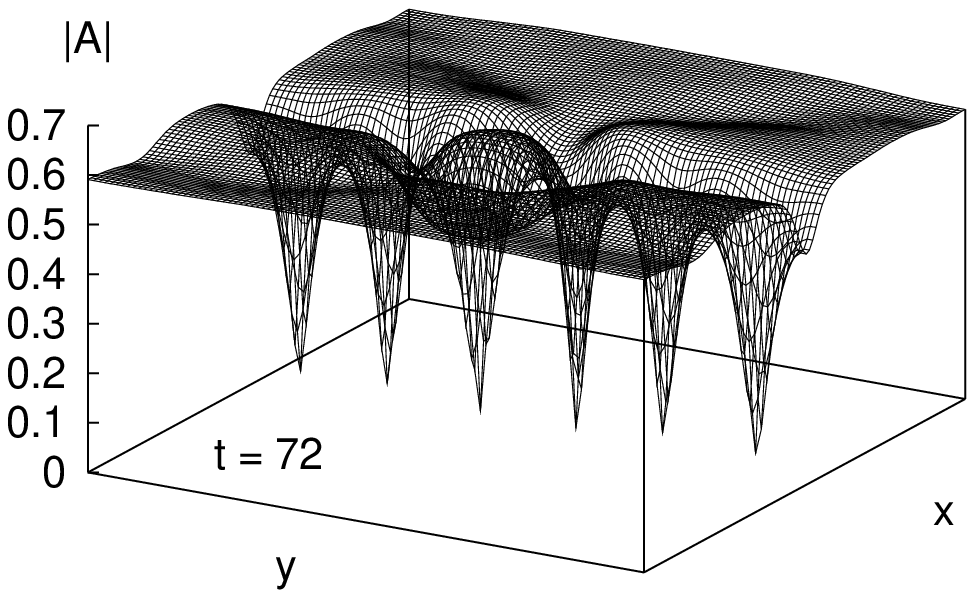}
\end{center}
\caption{Amplitude $|A|$ at time $t = 48$ and 72.}
\label{fig:amplitude evolution}
\end{figure}

The amplitude $|A|$ corresponding to this sequence is illustrated in
\fig{fig:amplitude evolution}.
At $t = 48$, the vortices do not quite look like vortex solitons, even though the number of vortices has been reduced to six.
Since the distance between adjacent vortices is still small, they overlap and are not quite independent.
The central hump around $x \sim y \sim 0$ is the remnant of the second annihilation event.
With time, the peaks tend to expand to the central region and are sufficiently apart to look and behave like vortex solitons.
By $t \sim 72$, we observe the formation of six solitonic structures as predicted in Ref.\cite{TBM-01} (see $t = 72$ in \fig{fig:amplitude evolution}).
Notice that the vortices disappear only when annihilated by another with opposite polarity.

\begin{figure} \begin{center}
	\includegraphics[width=12cm]{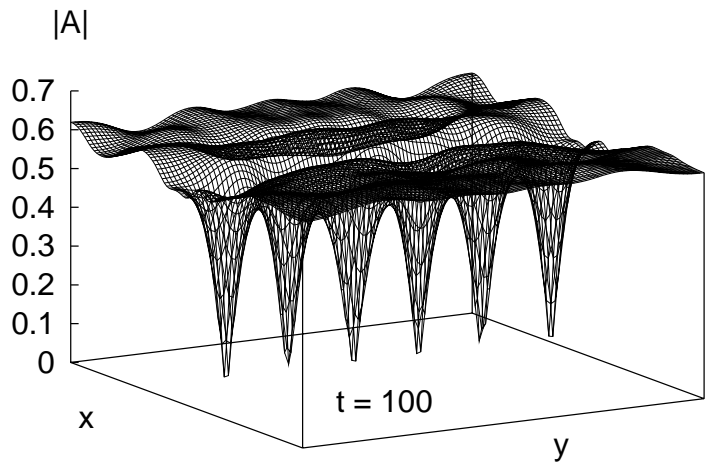}
\end{center}
\caption{Radiation appears after $t > 80$.}
\label{fig:radiation}
\end{figure}

After $t \gsim 80$, a propagating wave is clearly seen in the region $y < -10$ and $y > 10$ where the field was originally flat (see \fig{fig:radiation}).
This propagating field is the trace of the Cherenkov radiation which comes from the non-integrability of the system \cite{Iv-97}.
The radiation will propagate away from vortices.
In the finite calculation domain, the radiation will be reflected back at the boundary.

When we extend the time evolution further, we observe the third annihilation event around $t \sim 160$.
However, this annihilation may be an artifact of the finite size of the domain.
When we carry out the simulation in a domain larger than ($L = 28$) but with the same initial conditions, the third annihilation event takes place at a later time while the times of the first and the second annihilation events remain unchanged.
The third annihilation event is likely to be driven by the reflection of the Cherenkov radiation at the boundary.
Equivalently, it may be concluded that six vortex solitons are stable in our domain and may stay forever.
It is noted that the distance among vortices at the final stage approximately coincides with the inverse of the wave number with the maximum growth rate $k_{\rm m}$.

\section{Summary}
\label{sec:summary}

We have demonstrated the dynamics of angular momenta creation in a highly relativistic electron-positron plasmas subject to the passage of a strong pulse of electromagnetic fields.
The system is governed by a generalized nonlinear Schr\"odinger equation with a defocusing inverse square root type nonlinearity.
It turns out that the one-dimensional dark soliton stripe solutions of this equation, just like those of the standard nonlinear Schr\"odinger equation, are unstable to transverse perturbations.
By carrying out a linear analysis, we have found that there exists an instability window of transverse wave numbers for the system.
By a numerical simulation of the fully nonlinear equation, we have shown that the transverse instability will yield, after a few annihilation events, a well-separated chain of vortex solitons with alternating, singly-charged polarity or topological charge ($m = \pm 1$).
These singly-charged vortex solitons are topologically stable and do not disappear unless they collide with their compliments and annihilate.
The number of the created vortex solitons seem to be determined by the inverse of the wave number $k_{\rm m}$ with the maximum linear growth rate.
For the box size $L = 28$, the six vortex soliton state is found to be robust.

We have suggested a simple and plausible mechanism of angular momentum generation in the MeV epoch of the Universe.
Electromagnetism, operating through the versatile substrate of the electron-positron plasma, seems to readily generate highly interesting, long-lived objects which are capable of carrying large amounts of mass, energy and angular momentum.
Since an initial localization of mass, energy and angular momentum is precisely the seed that gravity needs for eventual structure-formation, electromagnetism may have provided a key element in the construction of the large-scale map of the observable Universe.

\begin{acknowledgments}

This work is supported in part by a Grant-in-Aid from the Japanese Ministry of Education, Culture, Sports, Science and Technology, No. 12780353.
The work of V.I.B. was partially supported by the ISTC grant G-663, and S.M.M.'s work was supported by the U.S. Department of Energy Contract No. DE-FG03-96ER-54346.
\end{acknowledgments}

\newpage


\newpage

\begin{center}
{\bf {\Large Figure Captions}}
\end{center}

\begin{enumerate}
	\item Dispersion relation for $\lambda = 0.2$, $0.5$, and $1$.
	\item Eigenfunctions for $\lambda = 0.5$ corresponding to $k=0.1$, $0.5$, and $0.7$.
	\item Zero lines of real and imaginary parts of the field at time $t = 2$.
	\item Crossing of zeros at time $t = 40$, 44, 48 and 72.
	\item Amplitude $|A|$ at time $t = 48$ and 72.
	\item Radiation appears after $t > 80$.
\end{enumerate}

\newpage


\newpage

\newpage

\newpage

\newpage

\newpage

\end{document}